\documentclass[aps,amsmath,amssymb,preprint,longbibliography]{revtex4-1}

\usepackage{graphicx}
\usepackage{dcolumn}
\usepackage{bm}
\usepackage{amsmath}
\usepackage{xcolor}
\usepackage[utf8]{inputenc}
\usepackage[T1]{fontenc}
\usepackage{mathptmx}
\usepackage{etoolbox}
\usepackage{caption}
\usepackage{subcaption}
\usepackage{soul}

\renewcommand\hl[1]{#1} %

\begin{document}

\title[Finding passive, reciprocal metasurfaces for arbitrary wave transformations]{\hl{Finding passive, reciprocal metasurfaces for arbitrary wave transformations}}
\author{K.O. Arnold}
    \email{koa201@exeter.ac.uk}
 \affiliation{School of Physics and Astronomy, University of Exeter, Stocker Road, Exeter, EX4 4QL, UK}
 \author{C. Hooper}
 \affiliation{School of Physics and Astronomy, University of Exeter, Stocker Road, Exeter, EX4 4QL, UK}
\author{J.G. Smith }
 \affiliation{School of Physics and Astronomy, University of Exeter, Stocker Road, Exeter, EX4 4QL, UK}
\author{N. Clow}
\affiliation{DSTL, Porton Down Salisbury, Wiltshire, SP4 0JQ, UK}
\author{A.P. Hibbins}
 \affiliation{School of Physics and Astronomy, University of Exeter, Stocker Road, Exeter, EX4 4QL, UK}
\author{J.R. Sambles}
 \affiliation{School of Physics and Astronomy, University of Exeter, Stocker Road, Exeter, EX4 4QL, UK}
\author{S.A.R. Horsley}
 \affiliation{School of Physics and Astronomy, University of Exeter, Stocker Road, Exeter, EX4 4QL, UK}

\date{\today}
%
%
\begin{abstract}
\hl{We give a general design method for finding the passive, reciprocal surface impedance tensor required to enact any wave transformation.  We do this through characterising the surface in terms of a tensorial surface impedance, showing that a large family of impedance distributions can be found that perform an identical wave transformation.  Even when the conditions of reciprocity and passivity are imposed, there still remain many solutions to the design problem.}  We exploit this as a design method for metasurfaces, giving two examples where the metasurface rotates the input polarization and reshapes the output field, showing we can parameterize the set of equivalent reciprocal metasurfaces in terms of a single complex parameter.  \hl{In addition, through allowing dissipation and gain within the response, the surface can have many different functionalities in the orthogonal polarization, opening up a new route for the design of multiplexed metasurfaces.}
\end{abstract}

\maketitle

%
%
\section{\label{sec:Introduction}Introduction}
\begin{figure}[h!]
    \centering
    \includegraphics[width=\textwidth]{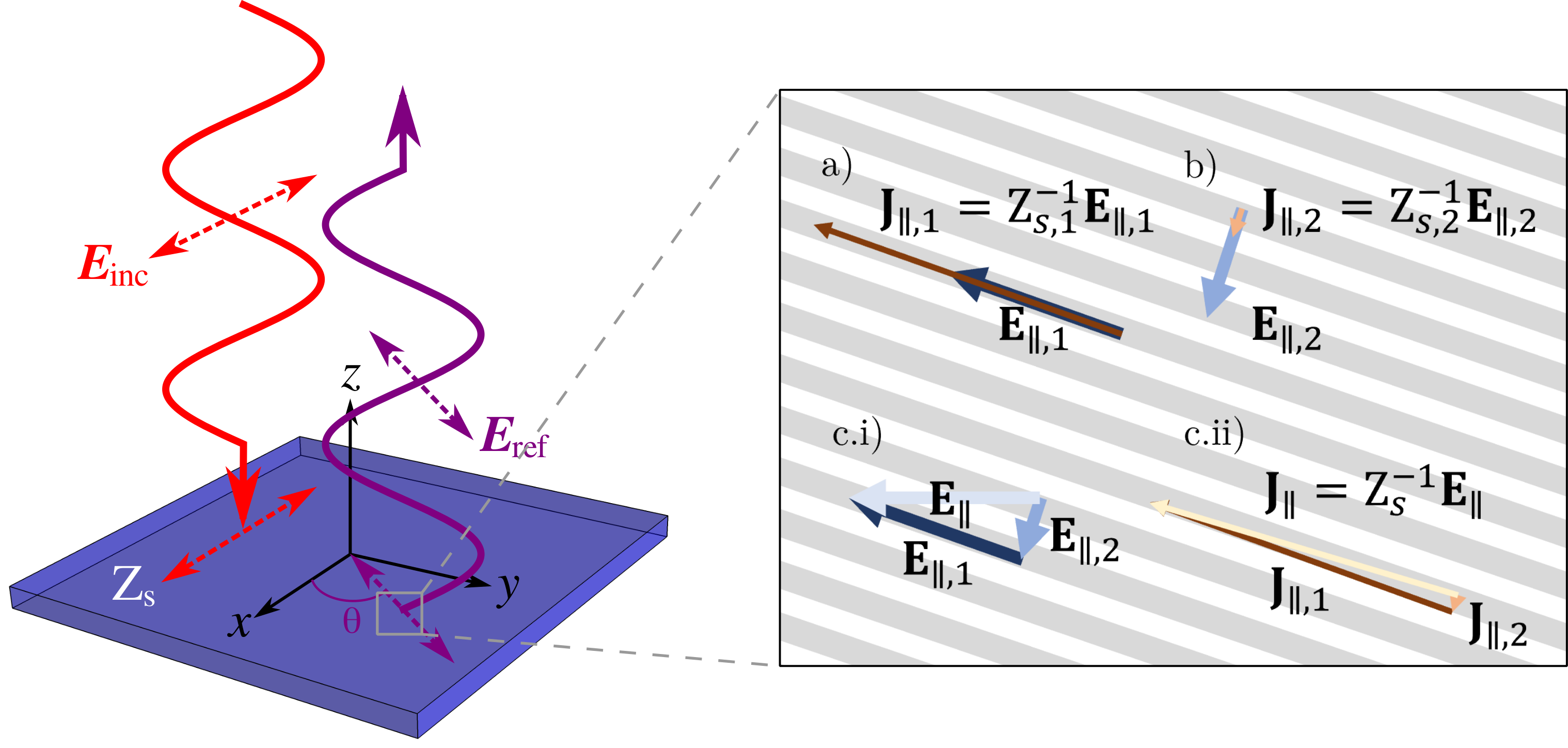}
    \caption{\textbf{Field transformations with tensorial metasurfaces:} A metasurface characterized in terms of a tensorial surface impedance, $\mathrm{Z}_{s}$ can perform an arbitrary transformation of an incident electromagnetic wave.  Here a normally incident wave with electric field aligned along the $x$--axis is transformed into a normally reflected wave with electric field at an angle $\theta$ to the $x$--axis, an example treated in Sec. \ref{sec:PC}.  The right hand panels (a,b) illustrate the behaviour of a tensorial metasurface in terms of the induced current $\mathbf{J}_{\parallel}$ and the electric field eigenvectors, $\mathbf{E}_{\parallel,i}$, in terms of which (ci,cii) any incident field can be decomposed}
    \label{fig: Rotate Pol}
\end{figure}

\hl{Metasurfaces are planar media containing an array of polarizable elements}~\cite{chen2016},\hl{carefully chosen to perform a desired transformation of an incoming wave.  Although this concept covers many areas of wave physics}~\cite{assouar2018},\hl{here we focus on metasurfaces for electromagnetic waves.  Within such surfaces, the elements are typically spaced closer than the wavelength, allowing the characterization of the array in terms of an effective electromagnetic boundary condition} \cite{achouri2021}.  \hl{Whereas partially or totally transmitting metasurfaces are typically characterized in terms of sheet transition conditions}~\cite{tretyakov2003,wu2018,lebbe2023}, \hl{or a polarizability}~\cite{holloway2009,zaluski2016},  \hl{impenetrable surfaces are usually characterized in terms of a surface impedance}~\cite{senior1995,tretyakov2003,maci2011}\hl{, as illustrated in Fig.}~\ref{fig: Rotate Pol}.  \hl{This amounts to specifying the ratio of the electric and magnetic fields at all points on the surface.  There are a variety of approaches to the design of such metasurfaces, controlling e.g. the local reflection phase over the surface}~\cite{capasso2014flat,yu2011light, chen2014high, wang2019multi}, \hl{or the coupling of surface waves to free--space radiation}~\cite{Sievenpiper2010holographic}.  \hl{In this work, we consider impenetrable metasurfaces described in terms of a tensorial surface impedance.  As we shall show, there is a non--uniqueness of the surface parameters with respect to a desired transformation of the field.  Besides being generally interesting, this can be exploited to impose the practical conditions of passivity and reciprocity on any metasurface design.  The greatest freedom occurs for non--reciprocal surfaces, but even reciprocal ones always contain a single free complex parameter.}
%
%

\hl{Although the terminology of a `metasurface' is little more than a decade old, the idea is a continuation of a large body of earlier work on frequency selective surfaces}~\cite{munk2005}, \hl{which was revived when similar ideas began to be implemented in optics}~\cite{yu2011light}.  \hl{Recent work has taken the metasurface idea significantly further forward.  For example, through matching the electric and magnetic polarizability of the surface elements in a transmitting sheet, all reflection can be eliminated, resulting in so--called `Huygens' metasurfaces}~\cite{pfeiffer2013,epstein2016,chen2018}, \hl{which sculpt the field while maintaining its direction of propagation. More recently, more exotic metasurfaces have been considered, including non--Hermitian and PT--symmetric elements}~\cite{fleury2014,tapar2021,fan2022}, \hl{which contain carefully designed distributions of loss and/or gain, allowing for an unusual control over e.g. the polarization of the output field.  Novel forms of surface excitations have also been discovered at junctions between metasurfaces}~\cite{horsley2014,bisharat2017}, \hl{and most recently non--reciprocal}~\cite{lawrence2018} \hl{and space--time varying surface parameters have been considered}~\cite{li2020,wang2020}, \hl{enabling both multiplexed functionality and spectral control.}

\hl{As mentioned above, here we investigate the uniqueness of the tensorial surface impedance, $\mathrm{Z}_{s}$, when enforcing a single given transformation of the electromagnetic field.  As well as exploiting this to find practical (reciprocal and passive) sets of parameters, this finding is also relevant to the problem of designing multiplexed---more specifically polarization multiplexed---metasurfaces, where two or more functionalities are imposed in the same design.  Unless the surface properties are e.g. re--configurable}~\cite{shirmanesh2020}, \hl{it is well known that this is a difficult design problem, and there have been numerous previous approaches}~\cite{capers2022,dai2023,murtaza2024}, \hl{some of which are summarized in Ref.}~\cite{xu2021}.
%
%
\section{\label{sec:TIBC}The Tensorial Impedance Boundary Condition}

\hl{In this section we review the mathematical description of impenetrable metasurfaces in terms of the impedance boundary condition.}  We can characterize the reflection of electromagnetic waves from any thin impenetrable metasurface in terms of an impedance boundary condition written in terms of a $2\times2$ tensor valued surface impedance~\cite{senior1995,tretyakov2003}, $\mathrm{Z}_s$.  To understand the origin of this boundary condition, consider Ohm's law within the metasurface relating the electric field parallel to a boundary, $\mathbf{E}_\parallel$, to the surface currents, $\mathbf{j}_s$, via the surface conductivity, $\sigma_s$. By relating the surface currents to the magnetic field via $\mathbf{j}_{s}=\hat{\mathbf{n}}\times\mathbf{H}_\parallel$ where $\hat{\mathbf{n}}$ is the surface normal, we have an equation relating the surface electric and magnetic fields,
\begin{equation}
    \label{eq: Ohms Law}
    \mathbf{j}_s = \sigma_s\cdot\mathbf{E}_\parallel = \hat{\mathbf{n}}\times\mathbf{H}_\parallel.
\end{equation}
In general, the surface conductivity is a $2\times 2$ tensor, indicating anisotropy, i.e. the vectorial surface current may not be directly proportional to the surface electric field.  For example, a surface composed of an array of thin parallel wires aligned with the $\hat{\mathbf{z}}$ direction would be characterized by the surface conductivity $\sigma_s=\sigma_{zz}\,\hat{\mathbf{z}}\otimes\hat{\mathbf{z}}$, indicating that current can only flow along the axis of the wires and only flows in response to the $\hat{\mathbf{z}}$ component of the electric field.  As a matter of convention, interfaces are characterized not in terms of the conductivity but in terms of the surface impedance, which is its inverse, $\mathrm{Z}_s=\sigma_{s}^{-1}$.  This yields the impedance boundary condition, 
\begin{equation}
    \label{eq: Impedance BC}
    \mathbf{E}_\parallel=\mathrm{Z}_s\cdot\left(\hat{\mathbf{n}}\times\mathbf{H}_\parallel\right).\
\end{equation}
Note that the component of the cycle averaged Poynting vector normal to the metasurface is given by $\langle S_{n}\rangle=\frac{1}{2}{\rm Re}[\hat{\mathbf{n}}\cdot(\mathbf{E}_{\parallel}\times\mathbf{H}_{\parallel}^{\star})]=-\frac{1}{2}{\rm Re}[\mathbf{E}_{\parallel}\cdot(\hat{\mathbf{n}}\times\mathbf{H}_{\parallel}^{\star})]$ and therefore the power locally entering the metasurface $\langle P_{n}\rangle=-\langle S_{n}\rangle$ is proportional to the Rayleigh quotient of the impedance matrix,
\begin{equation}
    \langle P_{n}\rangle=\frac{1}{2}{\rm Re}\left[\left(\hat{\mathbf{n}}\times\mathbf{H}_\parallel\right)^{\star}\cdot\mathrm{Z}_s\cdot\left(\hat{\mathbf{n}}\times\mathbf{H}_\parallel\right)\right].\label{eq:surface_power_flow}
\end{equation}
Given the assumption stated in the introduction, that the metasurface is impenetrable, a positive value for $\langle P_{n}\rangle$ indicates a dissipative response, whereas a negative value indicates amplification, or gain.  Given that a lossless metasurface will have $\langle P_{n}\rangle=0$ for all possible incident fields, lossless surfaces have an impedance tensor that is `${\rm i}$' times a Hermitian matrix (a so--called anti--Hermitian operator).  In the general case, we can decompose the impedance into Hermitian and anti--Hermitian parts, $Z_{s}=H_1+{\rm i}\,H_2$, where $H_{1,2}$ are Hermitian matrices.  The dissipated power is thus proportional to the Rayleigh quotient of $H_1$, implying that passive surfaces (i.e. those without gain) have a positive semi--definite form of $H_1$.

Another important constraint on the impedance tensor is provided by reciprocity.  Representing the freedom to interchange sources and detectors, this implies the symmetry of the electromagnetic Green function~\cite{novotny2012} $G(\boldsymbol{x}_{1},\boldsymbol{x}_{2})=G^{\rm T}(\boldsymbol{x}_{1},\boldsymbol{x}_{2})$, which is equivalent to the symmetry of the operators in Maxwell's equations.  Reciprocity thus requires the surface impedance to be a symmetric tensor, $\mathrm{Z}_{s}=\mathrm{Z}_{s}^{\rm T}$.
%
%
\section{\hl{Non--uniqueness of metasurfaces for performing field transformations: imposing reciprocity and passivity}\label{sec:field_transformations}}

\hl{Supposing we want a metasurface to perform a given transformation of the field upon reflection, we can easily find a set of values for the impedance through substituting the desired input and output fields into Eq.} (\ref{eq: Impedance BC}).  \hl{To do this, we write our specified complex valued surface fields in terms of their magnitude, e.g. $|E|=(\mathbf{E}_{\parallel} \cdot\mathbf{E}_{\parallel}^{\star})^{1/2}$ and ``direction'' $\hat{\mathbf{e}}=\mathbf{E}_{\parallel}/|E|$,}
\begin{align}
    \label{eq: Field unit vectors}
    \mathbf{E}_\parallel(x,y) &= |E(x,y)|\hat{\mathbf{e}}(x,y)\nonumber\\
    \mathbf{H}_\parallel(x,y) &= |H(x,y)|\hat{\mathbf{h}}(x,y)
\end{align}
We note that we can impose \emph{any} surface fields here, these being the sum of any incident and reflected fields with polarization, phase and amplitude varying arbitrarily across the surface.  Substituting this decomposition into the earlier expression for the impedance boundary condition (\ref{eq: Impedance BC}) we obtain the requirement
\begin{equation}
    \mathrm{Z}_{s}\cdot(\hat{\mathbf{n}}\times\hat{\mathbf{h}})=\frac{|E|}{|H|}\hat{\mathbf{e}}\label{eq:Z-required}
\end{equation}
which tells us that at each point on the surface, the impedance tensor must rotate and rescale the complex unit vector $\hat{\mathbf{n}}\times\hat{\mathbf{h}}$ so that it is parallel to the electric field unit vector, $\hat{\mathbf{e}}$.  When our desired field transformation has $\hat{\mathbf{e}}$ everywhere parallel to $\hat{\mathbf{n}}\times\hat{\mathbf{h}}_{\parallel}$, this can be achieved with an isotropic metasurface, characterized in terms of a scalar impedance, $\mathrm{Z}_s=|E|/|H|$.  This is always possible in e.g. a $2D$ geometry, where one field component always lies out of the plane and the other always in the plane.

Nevertheless in the general case, a spatially varying, tensor valued impedance is required to satisfy (\ref{eq:Z-required}).  A simple solution is $\mathrm{Z}_{s}=(|E|/|H|)\,\hat{\mathbf{e}}\otimes(\hat{\mathbf{n}}\times\hat{\mathbf{h}})^{\star}$, where `$\otimes$' is the tensor product defined as e.g. $(\mathbf{a}\otimes\mathbf{b})_{ij}=\left(\mathbf{a}\mathbf{b}^T\right)_{ij}=a_{i}b_{j}$.  However, as explained in e.g. Ref.~\cite{kwon2020}, this form of surface impedance is most often neither passive ($\mathrm{Z}_{s}+\mathrm{Z}_{s}^{\dagger}$ not positive semi--definite), nor reciprocal ($\mathrm{Z}_{s}\neq \mathrm{Z}_{s}^{\rm T}$).  This certainly represents a challenge for any practical realizations of these metasurface parameters, as both a power source and e.g. magnetic elements would have to be embedded within the surface!  Yet there is an additional freedom that we can exploit to bring the design closer to reality.  We can always perform a transformation $\mathrm{Z}_{s}\to \mathrm{Z}_{s}+\mathbf{v}\otimes\hat{\mathbf{h}}$, where $\mathbf{v}$ is an arbitrary vector, without affecting condition (\ref{eq:Z-required}).  This leaves us with an impedance tensor of the general form,
\begin{equation}
    \label{eq: Z with V}
    \mathrm{Z}_s(x,y) = \frac{|E|}{|H|}\left[\hat{\mathbf{e}}\otimes\left(\hat{\mathbf{n}}\times\hat{\mathbf{h}}\right)^{\star} + \mathbf{v}\otimes\hat{\mathbf{h}}\right].
\end{equation}
This surface impedance performs the desired arbitrary field transformation (\ref{eq:Z-required}) for all choices of the complex, position dependent vector, $\mathbf{v}(x,y)$.  As just discussed, generally these metasurface parameters are non--reciprocal and non--passive.  The remainder of this paper explores the freedom in specifying $\boldsymbol{v}$ in Eq. (\ref{eq: Z with V}), in particular it exploitation to find reciprocal, passive surface parameters.
%
%
\subsection{Reciprocal metasurfaces:\label{sec:reciprocal}}

To illustrate special cases of the impedance distribution defined in Eq. (\ref{eq: Z with V}), we adopt a particular complex basis to express the impedance tensor.  Here, we use the two complex basis vectors $\hat{\mathbf{h}}$ and $\hat{\mathbf{n}}\times\hat{\mathbf{h}}^{\star}$ and the dual basis, $\hat{\mathbf{h}}^{\star}$ and $\hat{\mathbf{n}}\times\hat{\mathbf{h}}$.  We use a bar to denote components in the dual basis.  With this choice the electric field direction on the surface is given by $\hat{\mathbf{e}}=e_1\,\hat{\mathbf{h}}+e_{2}\,(\hat{\mathbf{n}}\times\hat{\mathbf{h}})^{\star}=\bar{e}_{1}\,\hat{\mathbf{h}}^{\star}+\bar{e}_{2}\,(\hat{\mathbf{n}}\times\hat{\mathbf{h}})$ and the arbitrary vector is written as $\mathbf{v}=\alpha\,\hat{\mathbf{h}}+\beta\,(\hat{\mathbf{n}}\times\hat{\mathbf{h}})^{\star}=\bar{\alpha}\hat{\mathbf{h}}^{\star}+\bar{\beta}(\mathbf{n}\times\hat{\mathbf{h}})$.  In terms of these basis vectors the reciprocal counterpart to the impedance tensor (\ref{eq: Z with V}) is given by
\begin{multline}
    \mathrm{Z}_s(x,y)=\frac{|E|}{|H|}\bigg\{e_{1}(x,y)\left[\,\hat{\mathbf{h}}\otimes(\hat{\mathbf{n}}\times\hat{\mathbf{h}})^\star+\,(\hat{\mathbf{n}}\times\hat{\mathbf{h}})^\star\otimes\hat{\mathbf{h}}\right]+e_{2}(x,y)\,(\hat{\mathbf{n}}\times\hat{\mathbf{h}})^\star\otimes(\hat{\mathbf{n}}\times\hat{\mathbf{h}})^\star\\
    +\alpha(x,y)\,\hat{\mathbf{h}}\otimes\hat{\mathbf{h}}\bigg\}\\
    \label{eq:reciprocal_Z}
\end{multline}
where we have here ensured that the metasurface is reciprocal, $\mathrm{Z}_{s}=\mathrm{Z}_{s}^{\rm T}$, through taking
\begin{equation}
    \beta=e_1,\label{eq:reciprocity-condition}
\end{equation}
which determines a single component of the arbitrary vector, i.e. $\mathbf{v}\cdot(\hat{\mathbf{n}}\times\hat{\mathbf{h}})=\hat{\mathbf{e}}\cdot\hat{\mathbf{h}}^{\star}$.

\hl{From Eq.} (\ref{eq:reciprocal_Z}) \hl{we can see that, irrespective of the field transformation we seek to enact, through exploiting the non--uniqueness in} (\ref{eq: Z with V}), \hl{it is always possible to find a \emph{reciprocal} metasurface that will perform this transformation.  Indeed, we have an infinite family of such surfaces as we are still free to choose the complex valued parameter $\alpha$.  Ultimately the physical reason for this design freedom is that at every point we are imposing the response to a given surface field.  Of course this leaves the response to the orthogonal surface field unspecified.  The somewhat surprising fact is that while the requirement of reciprocity removes some of this redundancy, it is not enough to completely determine the form of the surface impedance.  In the next section we will impose the second practical constraint of passivity.}
%
%
\subsection{Passive metasurfaces:\label{sec:passive}}

As described in Sec. \ref{sec:TIBC}, for a metasurface to be passive it cannot amplify the incident field, and therefore the power entering the surface obeys $\langle P_{n}\rangle=-\langle S_n\rangle>0$.  In terms of the unit vectors (\ref{eq: Field unit vectors}) and the basis introduced in the previous section, this is equivalent to the condition,
\begin{equation}
    \langle P_{n}\rangle={\rm Re}\left[\hat{\mathbf{e}}\cdot(\hat{\mathbf{n}}\times\hat{\mathbf{h}}^{\star})\right]={\rm Re}[\bar{e}_{2}]\geq 0.\label{eq:passive-condition}
\end{equation}
\hl{We note that the requirement} (\ref{eq:passive-condition}) \hl{on the field transformation must be implemented at every point on the surface at the design stage, right from the outset: we cannot hope for a passive surface impedance if our \emph{design} fields require the metasurface to generate energy!}  However, even assuming Eq. (\ref{eq:passive-condition}) holds, it may still be that the impedance (\ref{eq:reciprocal_Z}) exhibits gain when probed with \emph{other} incident fields.

\begin{figure}[h!]
\includegraphics[width=\textwidth]{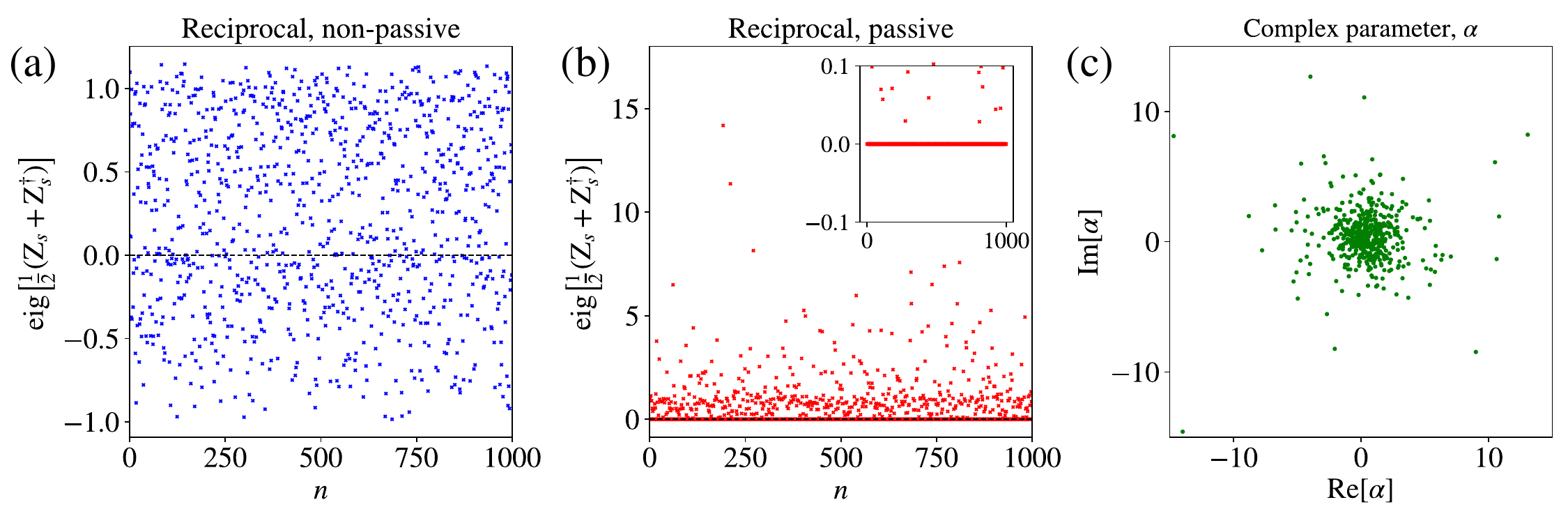}
\caption{\textbf{Randomly generated wave transformations:} Using a set of $500$ randomly generated complex unit vectors, $\hat{\mathbf{e}}$ and $\hat{\mathbf{h}}$, representing different choices of wave transformation, we computed the reciprocal surface impedance given by Eq. (\ref{eq:reciprocal_Z}) (with $|E|/|H|=1$, plus the constraint (\ref{eq:passive-condition})), and the eigenvalues of $(\mathrm{Z}_{s}+\mathrm{Z}_{s}^{\dagger})/2$, which determine the dissipation/gain in the metasurface. (a) $\alpha=0$ for all designs, giving rise to eigenvalues of both signs, generally requiring a non--passive metasurface. (b) $\alpha$ constrained by Eq. (\ref{eq:passive-condition-2}) (here taking the equality), giving rise to eigenvalues of a positive sign only, showing that all surfaces can be made passive, as desired. (c) Corresponding values of $\alpha$ for the metasurfaces considered in panel (b), showing that in general this parameter must be complex valued.\label{fig:random}}
\end{figure}

As explained in Sec. \ref{sec:TIBC}, to ensure the surface is passive, the Hermitian part of the impedance tensor must be positive semi--definite, i.e. for all complex unit vectors $\mathbf{u}$ we must have,
\begin{align}
    {\rm Re}\left[\mathbf{u}^{\star}\cdot \mathrm{Z}_{s} \cdot \mathbf{u}\right]&=\frac{|E|}{|H|}{\rm Re}\left[\bar{\alpha}|\bar{u}_{1}|^{2}+\bar{e}_{2}|\bar{u}_{2}|^2+\bar{e}_{1}\bar{u}_{1}^{\star}\bar{u}_{2}+\bar{\beta}\bar{u}_{2}^{\star}\bar{u}_{1}\right]\nonumber\\
    &\geq 0.\label{eq:passive_inequality}
\end{align}
We can ensure the real part of the quadratic form (\ref{eq:passive_inequality}) is always positive through imposing both condition (\ref{eq:passive-condition}) and choosing $\bar{\alpha}$, such that the determinant of the following Hermitian matrix is positive
\begin{equation}
    {\rm det}\left[\left(\begin{matrix}{\rm Re}[\bar{\alpha}]&\frac{\bar{\beta}+\bar{e}_{1}^{\star}}{2}\\\frac{\bar{\beta}^{\star}+\bar{e}_{1}}{2}&{\rm Re}[\bar{e}_{2}]\end{matrix}\right)\right]>0
\end{equation}
which implies the constraint
\begin{equation}
    {\rm Re}[\bar{\alpha}]\geq\frac{1}{{\rm Re}[\bar{e}_{2}]}\left|\frac{\bar{e}_{1}^{\star}+\bar{\beta}}{2}\right|^2\label{eq:passive-condition-2}.
\end{equation}
Fig. \ref{fig:random} uses a randomly generated set of surface electric and magnetic fields to show that imposing this constraint indeed results in a passive metasurface.  Note that if we impose zero dissipation in the response to the design field, ${\rm Re}[\bar{e}_{2}]=0$, the inequality (\ref{eq:passive-condition-2}) is generally impossible to fulfill.  This shows that reducing the efficiency of the design (i.e. \emph{decreasing} the reflected power) can help to find a set of metasurface parameters that are, overall passive.   We shall return to this point again in the next section.

Imposing conditions (\ref{eq:passive-condition}) and (\ref{eq:passive-condition-2}) therefore ensures that our metasurface design is gain--free.  If, on top of this, we also impose the constraint (\ref{eq:reciprocity-condition}), we find a surface that is both passive \emph{and} reciprocal.  However, there is not enough information in Eqns. (\ref{eq:reciprocity-condition}) and (\ref{eq:passive-condition-2}) to completely determine both complex parameters $\alpha$ and $\beta$.  There thus remains a non--uniqueness in the design of even reciprocal, passive metasurfaces.
%
%

\subsection{A special class of reciprocal, passive metasurfaces:\label{sec:special}}

A simple special case of the results given in Secs. \ref{sec:reciprocal} and \ref{sec:passive} is where the unit vector $\hat{\mathbf{h}}$ defined in Eq. (\ref{eq: Field unit vectors}) is real valued.  In this case, we no longer have to distinguish the vector basis from its dual, as defined in Sec. \ref{sec:reciprocal}, and barred and un--barred quantities become equal, i.e. $\bar{\alpha}=\alpha$, $\bar{\beta}=\beta$, $\bar{e}_{2}=e_{2}$, $\bar{e}_{1}=e_{1}$, etc.  Therefore, the three conditions---(\ref{eq:reciprocity-condition}), (\ref{eq:passive-condition}), and (\ref{eq:passive-condition-2})--for a passive, reciprocal metasurface become simply
\begin{align}
    \beta&=e_1\nonumber\\
    {\rm Re}[e_{2}]&\geq0\nonumber\\
    {\rm Re}[\alpha]&\geq\frac{1}{{\rm Re}[e_2]}|e_1|^2\label{eq:simple_constraints}
\end{align}
which, as discussed above still leaves ${\rm Im}[\alpha]$ undetermined, indicating an infinite family of passive reciprocal metasurfaces that perform the same transformation.

As an initial very simple example, consider a metasurface that completely absorbs incident waves when the electric field is aligned along the $x$ axis.  For a surface normal $\hat{\mathbf{n}}=\hat{\mathbf{z}}$, the direction of the electric field is $\hat{\mathbf{e}}=\hat{\mathbf{x}}$, and the magnetic field is $\hat{\mathbf{h}}=-\hat{\mathbf{y}}$, with $|E|/|H|=\eta_0$ where $\eta_0$ is the free space impedance.  In this case, $e_{1}=0$, $e_{2}=1$ and the reciprocal impedance (\ref{eq:reciprocal_Z}) reduces to
\begin{equation}
\mathrm{Z}_s=\eta_0\left(\hat{\mathbf{x}}\otimes\hat{\mathbf{x}}+\alpha\,\hat{\mathbf{y}}\otimes\hat{\mathbf{y}}\right).
\end{equation}
with all of the constraints in Eq. (\ref{eq:simple_constraints}) fulfilled so long as ${\rm Re}[\alpha]>0$.  Our design formulae therefore reproduces the expected result that a reciprocal, passive metasurface that perfectly absorbs $x$ oriented incident fields, must have the impedance component $\mathrm{Z}_{s,xx}$ equal to the free space impedance, and the orthogonal component, $\mathrm{Z}_{s,zz}$ must have a positive real part.  Note that the imaginary part of $\alpha$ is unconstrained.
\section{\label{sec:PC}Example: Polarization Conversion}

\hl{As a second example, we now take a slightly more complex illustration of the family of reciprocal metasurfaces described by Eq.} (\ref{eq:reciprocal_Z})\hl{, although we retain the assumption of surface uniformity: we consider a polarization converting metasurface.} A normally incident plane wave polarized along the $x$ axis impinges on a surface with surface normal $\hat{\mathbf{n}}=\hat{\mathbf{z}}$.  The desired field is reflected with some real amplitude $\rho$, with a $\theta$ change in the polarization angle away from $\hat{\mathbf{x}}$, as illustrated in Fig. \ref{fig: Rotate Pol}.  The surface electric field equals,
%
%
\begin{figure}[h!]
    \includegraphics[width=0.8\textwidth]{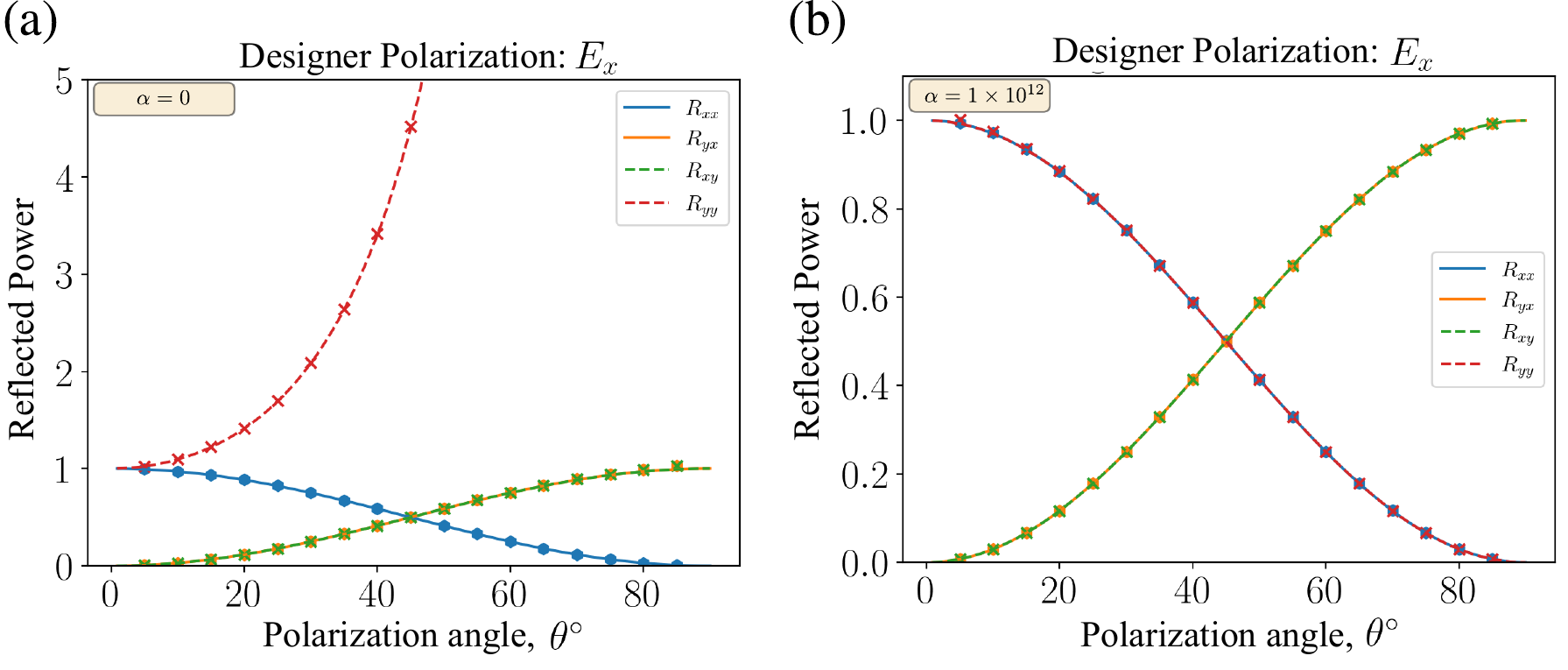}
    \caption{\textbf{Reflected power vs. polarization rotation angle $\theta$}: The co-- and cross--polarized reflectivities, $R_{ij}=|r_{ij}|^2$ (see appendix \ref{app:reflection}) for different metasurfaces, as a function of the polarization rotation angle $\theta$ from Fig. \ref{fig: Rotate Pol}.  (a) $\alpha=0$ and $\rho=1$.  As anticipated from Fig. \ref{fig:random}, without constraining the value of $\alpha$, the designer polarization ($R_{xx}$ and $R_{yx}$) conserves power, but the secondary polarization ($R_{yy}$ and $R_{xy}$) is amplified. (b) As in panel (a), but for $\alpha=1\times10^{12}$, showing that the surface approaches being passive for large real $\alpha$, as predicted by Eq. (\ref{eq:alpha_constraint_example})}
    \label{fig: Ref vs Ang a=0}
\end{figure}

\begin{align}
    \label{eq: Field at surface}
    \mathbf{E} &= \left[{\rm e}^{-\mathrm{i}k_0z}+r\cos{(\theta)}{\rm e}^{\mathrm{i}k_0z}\right]\hat{\mathbf{x}}+\rho\sin(\theta){\rm e}^{\mathrm{i}k_0z}\,\hat{\mathbf{y}}\nonumber\\[10pt]
    &\stackrel{z=0}{\to} \hat{\mathbf{x}}
+\rho\hat{\mathbf{m}},
\end{align}
where we have introduced the unit vector, $\hat{\mathbf{m}}=\cos(\theta)\hat{\mathbf{x}}+\sin(\theta)\hat{\mathbf{y}}$.  The previous sections show that the surface impedance enacting this transformation is characterized in terms of the electric and magnetic field unit vectors $\hat{\mathbf{e}}$ and $\hat{\mathbf{h}}$, which in this case are given by,
\begin{align}
    \hat{\mathbf{e}}&=\frac{\hat{\mathbf{x}}+\rho\hat{\mathbf{m}}}{\sqrt{1+\rho^2+2\rho\cos(\theta)}}\nonumber\\[10pt]
    \hat{\mathbf{h}}&=\frac{-\hat{\mathbf{y}}+\rho\hat{\mathbf{z}}\times\hat{\mathbf{m}}}{\sqrt{1+\rho^2-2\rho\cos(\theta)}}.\label{eq:example_unit_vectors}
\end{align}
In this example, the unit vectors (\ref{eq:example_unit_vectors}) describing the field on the surface are purely real.  Therefore, we can find a passive reciprocal metasurface using the results derived in Sec. \ref{sec:special}.  The reciprocal surface impedance tensor is given by Eq. (\ref{eq:reciprocal_Z}), with the vector components $e_1=\hat{\mathbf{e}}\cdot\hat{\mathbf{h}}$ and $e_2=\hat{\mathbf{e}}\cdot(\hat{\mathbf{n}}\times\hat{\mathbf{h}})$ equal to,
\begin{align}
    e_1&=-\frac{2\rho\sin(\theta)}{\sqrt{(1+\rho^2)^2-4\rho^2\cos^2(\theta)}}\nonumber\\
    e_2&=\frac{1-\rho^2}{\sqrt{(1+\rho^2)^2-4\rho^2\cos^2(\theta)}}
\end{align}
In terms of these two components, the constraints for a passive metasurface (\ref{eq:simple_constraints}) are fulfilled if both $\rho^2\leq 1$ (i.e. the reflected wave does not contain a greater energy flux than the incident one) and,
\begin{equation}
    {\rm Re}[\alpha]\geq\frac{1}{1-\rho^2}\frac{4\rho^2\sin^2(\theta)}{\sqrt{(1+\rho^2)^2-4\rho^2\cos^2(\theta)}}\label{eq:alpha_constraint_example}
\end{equation}
with the imaginary part of $\alpha$ free.

\begin{figure}[h!]
    \centering
    \includegraphics[width=0.5\textwidth]{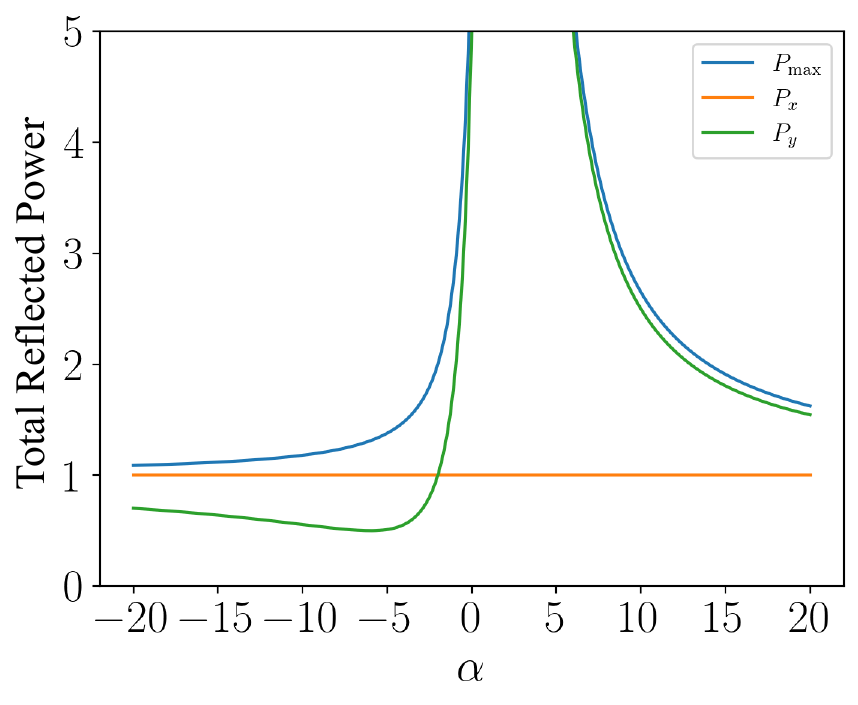}
    \caption{\textbf{Polarization multiplexed functionality}: for a $\theta=\pi/4$ polarization rotator (the design field incident with $x$ polarization), incident $y$ polarized waves can be transformed very differently, depending on the value of the complex parameter $\alpha$ in Eq. (\ref{eq:reciprocal_Z}) for a reciprocal surface impedance.  The above shows that, while the proportional of reflected power $P_x$ for $x$ polarized incidence is always unity, that for orthogonal incidence, $P_y$ is subject to either absorption or amplification, depending on the sign of ${\rm Re}[\alpha]$, with lasing--like behaviour at $\alpha=2$.  Note that the maximum output power, $P_{\rm max}$, here found through an appropriate combination of $x$ and $y$ polarizations always exhibits gain except as $|\alpha|\to\infty$.  This is consistent with Fig. \ref{fig: Ref vs Ang a=0} and constraint (\ref{eq:alpha_constraint_example}), as $\rho=1$, $\alpha$ must become infinite for the surface to be passive.}
    \label{fig: Pmax vs Alpha}
\end{figure}

Given the uniformity of the surface impedance in this case, we can analytically find the reflection from the surface defined by Eqns. (\ref{eq:reciprocal_Z}), (\ref{eq:example_unit_vectors}), and (\ref{eq: Field at surface}) to verify the above predictions.  The reflection amplitudes, $r_{xx}$, $r_{xy}$, $r_{yx}$, and $r_{yy}$ are defined such that the first index corresponds to the output polarization, and the second index to the input polarization.  The expressions for these are given in appendix \ref{app:reflection}. The power carried in each polarization channel is given by the square of these amplitudes, e.g. $R_{xx}=|r_{xx}|^2$ tells us the fraction of power reflected in the $x$ channel for incidence in the same channel. Fig. \ref{fig: Ref vs Ang a=0} shows both the results of a full wave simulation, and the reflection computed analytically, verifying both the reciprocity of the surface, $R_{xy}=R_{yx}$ and that incident $x$ polarized waves are reflected with the desired polarization. As shown in Fig. \ref{fig: Ref vs Ang a=0}a, when $\alpha=0$, at designer incidence, the output power is conserved. Meanwhile for the secondary incident polarization, the surface reflects with increased power, even a divergent value when $\alpha=2$, shown in Fig. \ref{fig: Pmax vs Alpha}.  For the surface to be passive (as in e.g. \cite{chen2014polarization,Cheng2014polarization,feng2013polarization,huang2014polarization,Yan2005polarization,Yang2014polarization,Zhu2013polarization}), our constraint (\ref{eq:alpha_constraint_example}) shows that for $\rho=1$ (i.e. $100\%$ efficiency for $x$ polarized incidence), this is only possible when ${\rm Re}[\alpha]\to\infty$.  Figs. \ref{fig: Ref vs Ang a=0}b and \ref{fig: Pmax vs Alpha} show this is indeed the case, with the amplification diminishing with increasing $\alpha$.





%
%

%
%
\begin{figure}[h!]
    \centering
    \includegraphics[width=0.9\textwidth]{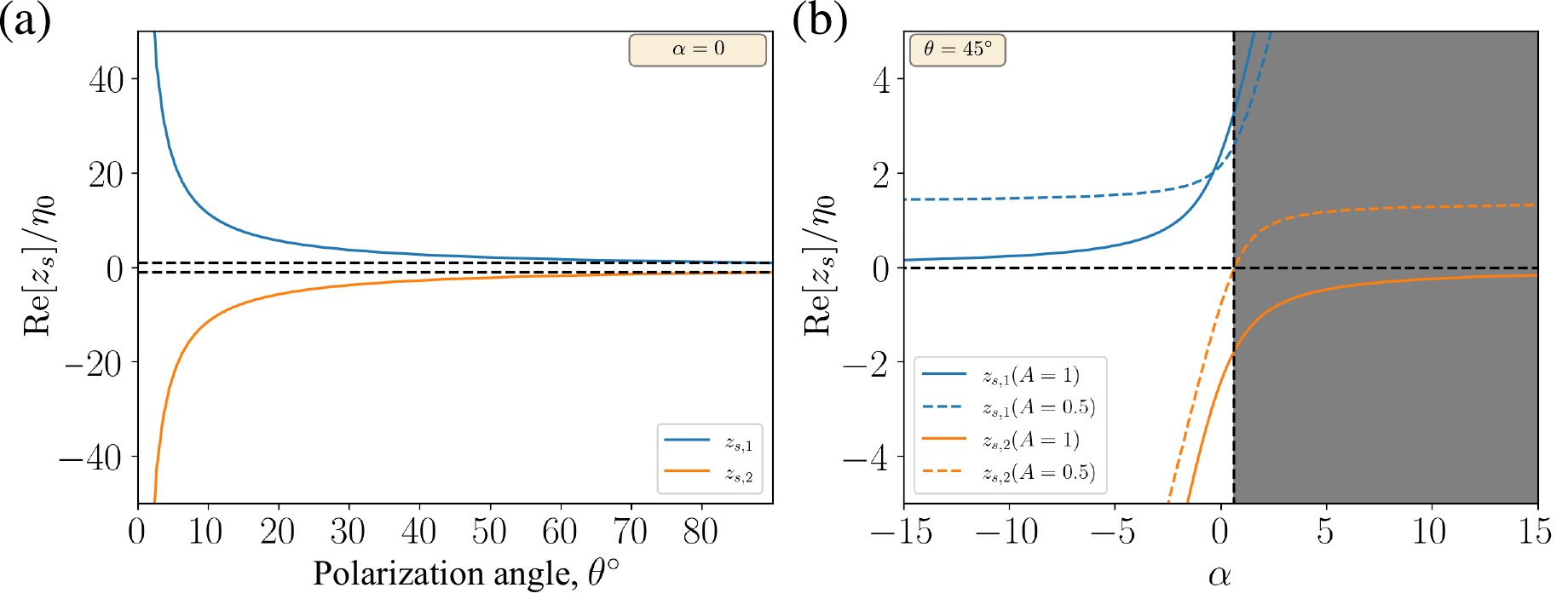}
    \caption{\textbf{Real part of $\mathrm{Z}_{s}$ eigenvalues for varying rotation angle, $\theta$ and design parameter, $\alpha$:} (a) Changing reflected polarization angle, $\theta$, at $\alpha=0$. As already indicated in Fig. \ref{fig: Ref vs Ang a=0}, one eigenvalue always corresponds to loss and the other to gain within these metasurfaces.  Note that as $\theta\rightarrow90^\circ$, the eigenvalues tends towards $\pm\eta_0$ (i.e. a perfect absorber for the incident polarization and a infinite gain surface for the reflected polarization), denoted as black dashed lines.  Note also that at $\theta=0$, both eigenvalues diverge.  This does not indicate infinite gain, as such a large impedance mismatch with free space only leads to total reflection. (b) Eigenvalues for $100\%$ ($\rho=1$) and $50\%$ ($\rho=0.5$) efficient reflectors, as a function of $\alpha$.  The dashed line shows the constraint (\ref{eq:alpha_constraint_example}) where the surface is passive, $\alpha>0.646$, above which the eigenvalues have a positive real part.}
    \label{fig: eig vs alpha}
\end{figure}

To demonstrate the validity of our inequality (\ref{eq:alpha_constraint_example}), we work in terms of the eigenvectors of the surface impedance tensor: those directions along which an applied electric field results in a parallel current density.   As discussed in Sec. \ref{sec:TIBC}, the power dissipated within the surface when the surface current $\hat{\mathbf{n}}\times\hat{\mathbf{H}}_{\parallel}$ is parallel to one of these eigenvectors $\mathbf{u}_{i}$ is, from Eq. (\ref{eq:surface_power_flow}),
\begin{align}
    \label{eq: Orthogonal Powers}
    \langle P_{n,i}\rangle&=\frac{1}{2}{\rm Re}\left[\mathbf{u}_{i}^{\star}\cdot\mathrm{Z}_{s}\cdot\mathbf{u}_{i}\right]\nonumber\\
    &=\frac{1}{2}\left|\mathbf{u}_{i}\right|^{2}{\rm Re}[z_{s,i}].
\end{align}
where \(\mathrm{Z}_{s}\mathbf{u}_{i} = z_{s,i}\mathbf{u}_{i}\). Therefore if \({\rm Re}[z_{s,i}] < 0\) for either \(i = 1\) or \(i = 2\), then the surface cannot be passive. However this is only a necessary condition.  In general, the dissipated power should be positive for a general superposition, $c_1\mathbf{u}_{1}+c_2\mathbf{u}_{2}$. This power equals:
\begin{align}
    \label{eq: General Powers}
    \langle P_{n}\rangle = |c_1|^2 \langle P_{n,1}\rangle + |c_2|^2 \langle P_{n,2}\rangle + \frac{1}{2}{\rm Re}\left[z_{s,1}\,c_{1}c_{2}^{\star}\mathbf{u}_{2}^{*} \cdot \mathbf{u}_{1} + z_{s,2}\,c_{1}^{\star}c_{2}\mathbf{u}_{1}^{*} \cdot \mathbf{u}_{2} \right] \geq 0 .
\end{align}
When the two eigenvectors are orthogonal, this reduces to \(\langle P_{n}\rangle = |c_1|^2\langle P_{1}\rangle + |c_2|^2\langle P_{2}\rangle>0\), which leads to the requirement that the eigenvalues of $\mathrm{Z}_{s}$ should have a positive real part.  In the current example, the surface impedance matrix is both real and symmetric (provided the free parameter $\alpha$ is real).  Therefore its eigenvectors are orthogonal and we simply require both eigenvalues $z_{s,1}$, $z_{s,2}$ to have a positive real part.  Fig. \ref{fig: eig vs alpha} shows the real part of the two eigenvalues of the impedance tensor (\ref{eq:reciprocal_Z}) both as a function of the angle of polarization rotation and the parameter $\alpha$.   In Fig. \ref{fig: eig vs alpha} we see that the surface indeed becomes passive once the inequality (\ref{eq:alpha_constraint_example}) is satisfied.

\section{Example: Conversion of a cylindrical TE--polarized wave into a TM--polarized plane wave}\label{Sec:Lc_to_pw}

\begin{figure}
    \centering
    \includegraphics[width=0.8\linewidth]{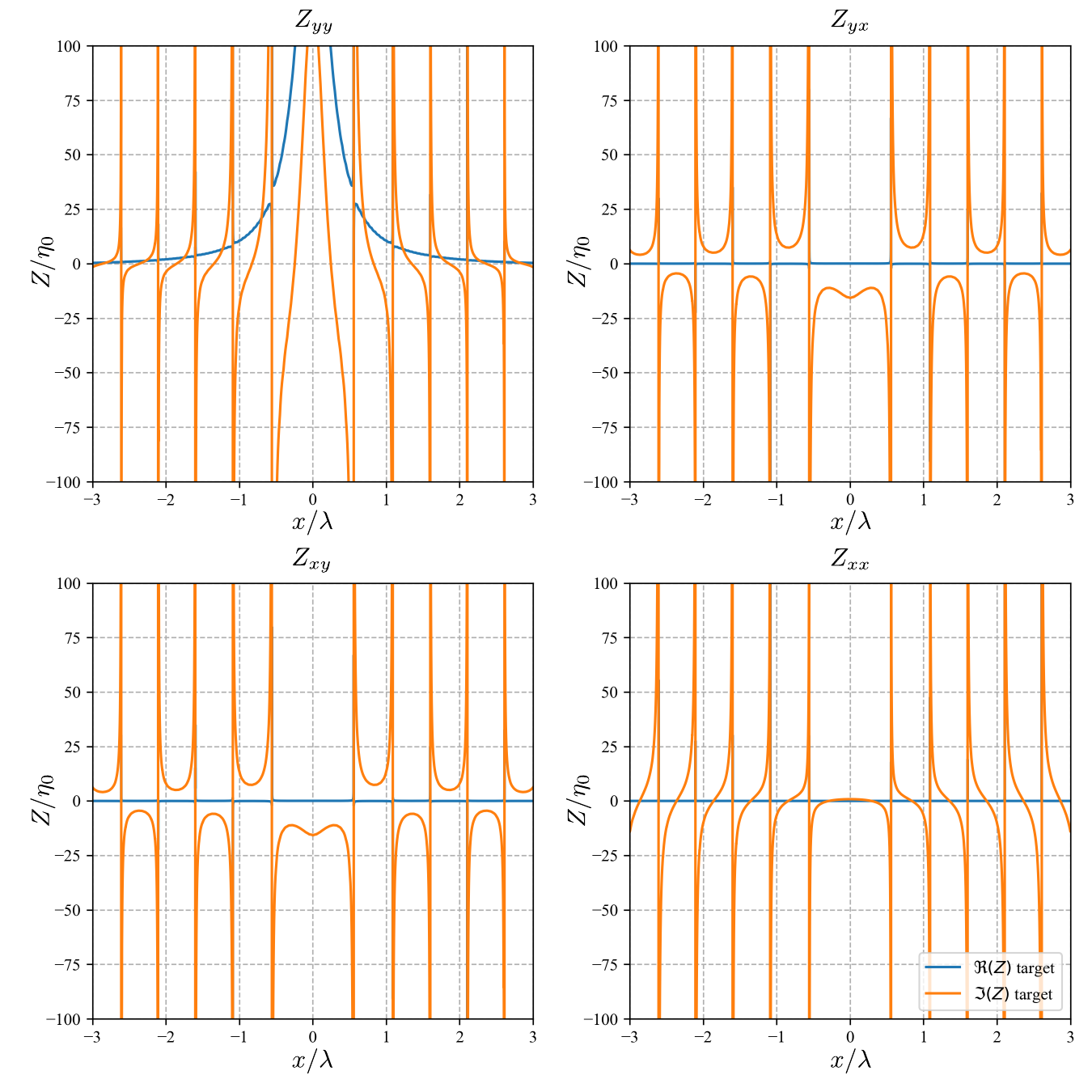}
    \caption{\hl{\textbf{Impedance tensor distribution for cylindrical to plane wave conversion:}  The constant amplitude of the reflected plane wave is a factor of 0.09 times the maximum of the incident field at the impedance boundary.}}
    \label{fig:lcimp}
\end{figure}

In the previous section we showed the simplest possible example: a normal incidence plane wave source.  We now demonstrate the applicability of our method to more general, inhomogeneous wave transformations. We use it to calculate the impedance profile required to convert an incident cylindrical wave into a plane wave.   To calculate the full tensorial impedance required at the surface, we numerically modelled a line curent in free space (Transverse Electric (TE) polarization) at a distance of $\lambda/4$ from the interface and tabulated the surface field as shown in appendix~\ref{app:Line current fields}.  The reflected field was chosen to be a Transverse Magnetic (TM) polarised plane wave. The sum of these two surface fields were then used as the fields in our design formula, Eq.(\ref{eq:reciprocal_Z}).  The components of the resulting impedance tensor, $\boldsymbol{Z}$ are shown in Fig. \ref{fig:lcimp}, with the distribution of the control parameter $\alpha$ used to calculate this impedance is given in appendix \ref{app:Passivealpha}. As expected the distribution is symmetric about the value of $x=0$ (the transverse position of the line current source), and oscillates as required to convert the incident field into radiation directed along the surface normal.

\begin{figure}
    \centering
    \includegraphics[width=0.8\linewidth]{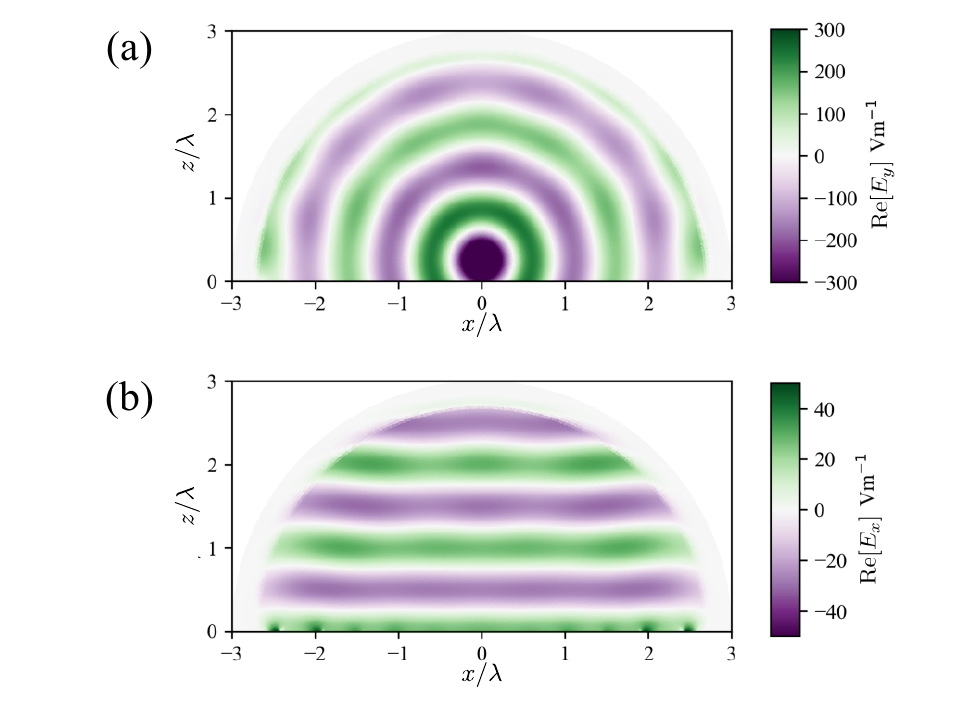}
    \caption{\hl{\textbf{Numerical verification of cylindrical to plane wave conversion:}  (a) The $\boldsymbol{\hat{y}}$ polarised electric field excited by the line current. (b) The $\boldsymbol{\hat{x}}$ polarized plane wave resulting from the reflection of the incident field from the impedance boundary.}}
    \label{fig:fieldplots}
\end{figure}

Fig. \ref{fig:fieldplots} shows a numerical simulation of the out of plane and in plane components of the electric field for a line current source placed a distance $\lambda/4$ above the impedance distribution shown in Fig. \ref{fig:lcimp}.  In Fig. \ref{fig:fieldplots} (a) we plot the $E_y$ electric field component, showing the expected field generated by the line current source.  Fig. \ref{fig:fieldplots} (b) shows the $E_x$ field component, from which we can identify the desired plane wave propagating normal to the impedance boundary. Note that in Fig. \ref{fig:fieldplots} (b) we can observe large magnitude features in the fields at the surface that are not part of the desired output field.  These maxima are features that depend on the mesh used in the solver and occur where the determinant of the impedance matrix changes sign.
This example shows the applicability of our method to find passive, tensorial impedance distributions that convert between any specified incident waveform and any desired reflected one.

%
%
\section{Realization of the metasurface design}

\begin{figure}
    \centering
    \includegraphics[width=1.0\linewidth]{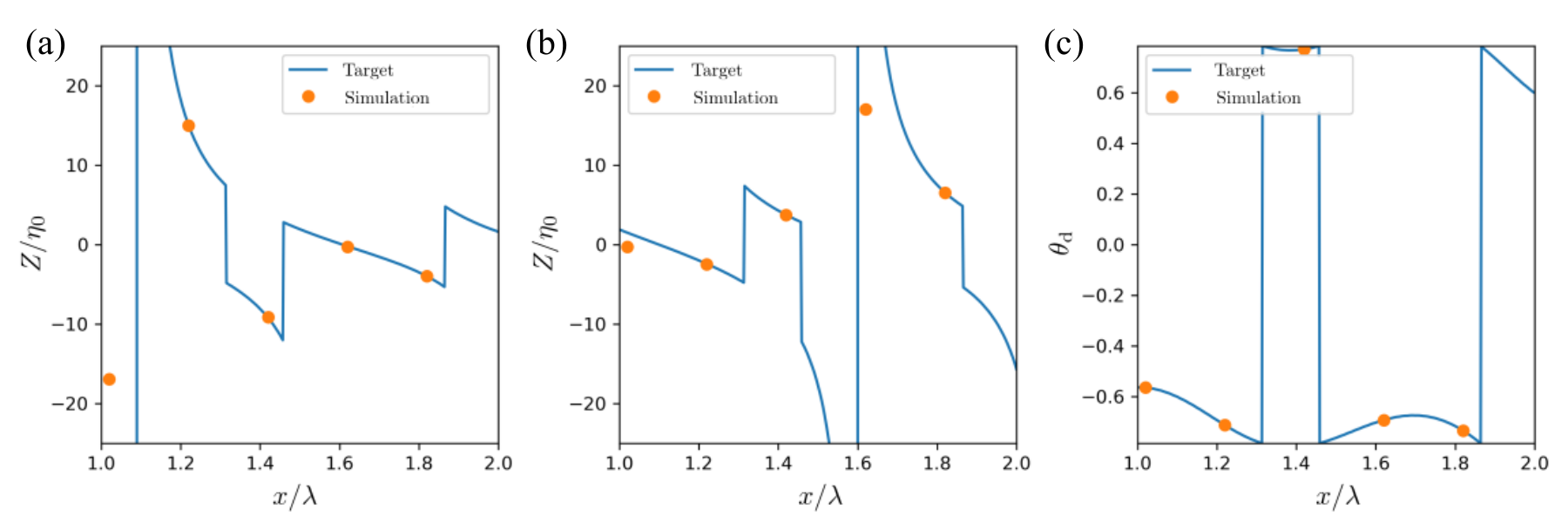}
    \caption{\hl{\textbf{Example realization of impedance profile:} Using the elliptical patch unit cell shown in Appendix}~\ref{ap:realization}, \hl{we calculated the eigenvalues, $Z_{xx}'$ and $Z_{yy}'$ (panels a and b), of the impedance tensor (dots), varying the dimensions of the patch to fit the desired distribution of eigenvalues calculated from Eq.} (\ref{eq:reciprocal_Z}) \hl{shown in Fig.} (\ref{fig:lcimp} \hl{(solid line).  Note the jumps in eigenvalues in panels (a) and (b) simply correspond to the ordering given by the numerical linear algebra routine and do not affect the form of the impedance tensor (the corresponding jump in angle in panel c corrects this change of order).  To reproduce the tensor we then determined the rotation angle $\theta_d$ necessary to diagonalize the impedance tensor (panel c), thus determining the rotation of each patch on the surface.} }
    \label{fig:diagfit}
\end{figure}

\hl{The above formalism shows how we can obtain a passive, reciprocal tensorial impedance distribution to realise any given wave transformation.  However, in most cases we cannot directly implement an impedance distribution.  We must find a unit cell of the metasurface that gives an adequate approximation to the desired impedance.  Here we give an example of the procedure that can be used to find an implementation.  We take a discrete set of points in the $\lambda \leq x \leq 2\lambda$ section of the impedance distribution shown in Fig.} \ref{fig:lcimp}.  \hl{Taking an implementation as an elliptical patch on dielectric layer, tuned near the half wave resonance (see appendix D), we then diagonalise the impedance distribution shown in Fig. }\ref{fig:lcimp} \hl{by applying a rotation matrix (i.e. choosing the angle of the ellipse as a function of position on the surface).  The diagonalised distribution is shown as the blue ``target'' lines in Fig. }\ref{fig:diagfit}, \hl{and the dimensions of the elliptical patch are varied and related to the surface impedance at a given set of points on the surface as described in e.g.}\cite{PatelModAn,smith2024method}, \hl{the results are here shown as the dots in Fig.} \ref{fig:diagfit}.  \hl{To obtain the full impedance, each patch is then rotated using the same matrix used to diagonalize the designed impedance matrix.

Note that here we have chosen to ignore the real part of the impedance, which is small in this part of the distribution.  To implement the anisotropic loss required by the full design we can e.g. include a set of lossy conducting wires, or anisotropic dielectric in the unit cell.  Connected to this, in many cases the real and imaginary parts of the tensorial surface impedance may not be simultaneously diagonalized using a rotation of the unit cell.  In these cases we must use a unit cell containing e.g. dielectrics or stacked conducting patches with different principal axes, rotating their relative angle and the angle of the full unit cell to obtain the desired tensorial impedance distribution.}

\section{Summary and Conclusions}

Impenetrable metasurfaces are often characterized in terms of a scalar valued surface impedance, the value of which encodes both the phase and amplitude of the reflected field.  However, as has been previously recognised, the use of a tensorial surface impedance---practically implemented via an array of anisotropic, polarizable elements---offers a much richer set of possible field transformations.  Here we have exploited the non-uniqueness of the metasurface parameters to solve the problem of designing tensorial metasurfaces that enact an arbitrary field transformation while remaining reciprocal, passive, or both.

Our first finding is that it is always possible to find a reciprocal metasurface, whatever the wave transformation we wish to perform.  The tensorial surface impedance is given by Eq. (\ref{eq:reciprocal_Z}) and there are an infinite number of such surfaces, parameterized in terms of a single complex number, $\alpha$.  In general this family of metasurfaces will exhibit gain and loss for some incident fields, showing that through carefully implementing loss and/or gain within the surface, we may obtain multiplexed functionality.  As an example, we showed in Sec. \ref{sec:PC} that a metasurface rotating $x$ polarized fields by an angle $\theta$ could simultaneously show a variety of different functionalities for $y$--polarized radiation, e.g. infinite amplification, absorption, or rotation.

Nevertheless, despite a large body of fascinating work (e.g. \cite{fleury2014}) and recent experimental progress~\cite{tapar2021,fan2022}, metasurfaces with gain are much more difficult to implement than lossy ones.  Our second finding is that, provided the complex parameter $\alpha$ satisfies the inequality (\ref{eq:passive-condition-2}), we can simultaneously ensure the metasurface is reciprocal and passive (without gain) for arbitrary incident and scattered fields.  Surprisingly the surface parameters remain non--unique, and there remain an infinite family of surfaces that perform the same transformation.  One important point is that the right hand side of the inequality (\ref{eq:passive-condition-2}) diverges when the initial design is made $100\%$ efficient (zero dissipation, all energy reflected).  This actually reduces the design freedom, implying that the free parameter $\alpha$ must be taken to infinity in order that the surface be both lossless and reciprocal. Therefore, when the efficiency of the design is reduced (e.g. reducing the overall amplitude of the reflected field), we find that this \emph{increases} the range of reciprocal, passive metasurfaces that can enact the transformation.  This was demonstrated in two examples of polarization rotating metasurface, where e.g. Fig. \ref{fig: eig vs alpha}b shows that reducing the efficiency immediately yields an increased range of passive reciprocal surfaces that would otherwise require ${\rm Re}[\alpha]\to\infty$.   We have thus found that a careful choice of loss within the metasurface actually aids the design, allowing for polarization multiplexed functionality, at the price of reducing the efficiency.

%
%
\appendix
\section{Reflection from a tensor impedance surface\label{app:reflection}}
Writing the components of the surface impedance tensor in terms of the Cartesian basis vectors, $\hat{\mathbf{x}}$and $\hat{\mathbf{y}}$ as,
\begin{equation}
    \mathrm{Z}_{s}=\left(\begin{matrix}Z_{xx}&Z_{xy}\\Z_{yx}&Z_{yy}\end{matrix}\right)
\end{equation}
and assuming a wave at normal incidence with electric and magnetic fields given by ($A_x$ amplitude incident with $x$ polarization and $A_y$ amplitude incident with $y$ polarization),
\begin{align}
\mathbf{E}&=A_{x}\left[{\hat{\mathbf{x}}}\left({\rm e}^{-{\rm i}k_0 z}+r_{xx}{\rm e}^{{\rm i}k_0 z}\right)+{\hat{\mathbf{y}}}r_{yx}{\rm e}^{{\rm i}k_0 z}\right]+A_{y}\left[{\hat{\mathbf{y}}}\left({\rm e}^{-{\rm i}k_0 z}+r_{yy}{\rm e}^{{\rm i}k_0 z}\right)+{\hat{\mathbf{x}}}r_{xy}{\rm e}^{{\rm i}k_0 z}\right]\nonumber\\
\eta_0\mathbf{H}&=A_{x}\left[{\hat{\mathbf{y}}}\left(-{\rm e}^{-{\rm i}k_0 z}+r_{xx}{\rm e}^{{\rm i}k_0 z}\right)-{\hat{\mathbf{x}}}r_{yx}{\rm e}^{{\rm i}k_0 z}\right]+A_{y}\left[{\hat{\mathbf{x}}}\left({\rm e}^{-{\rm i}k_0 z}-r_{yy}{\rm e}^{{\rm i}k_0 z}\right)+{\hat{\mathbf{y}}}r_{xy}{\rm e}^{{\rm i}k_0 z}\right],
\end{align}
the impedance boundary condition defined above in Eq. (\ref{eq: Impedance BC}) leads to the following expressions for the four reflections coefficients $r_{ij}$,
\begin{align}
    r_{xx}&=\frac{(Z_{xx}-1)(Z_{yy}+1)-Z_{xy}Z_{yx}}{(Z_{xx}+1)(Z_{yy}+1)-Z_{xy}Z_{yx}}\\
    r_{xy}&=\frac{2Z_{yx}}{(Z_{xx}+1)(Z_{yy}+1)-Z_{xy}Z_{yx}}\\
    r_{yx}&=\frac{2Z_{xy}}{(Z_{xx}+1)(Z_{yy}+1)-Z_{xy}Z_{yx}}\\
    r_{yy}&=\frac{(Z_{xx}-1)(Z_{yy}-1)-Z_{xy}Z_{yx}}{(Z_{xx}+1)(Z_{yy}+1)-Z_{xy}Z_{yx}}.
\end{align}
These expressions we used to generate Figs. \ref{fig: Ref vs Ang a=0}--\ref{fig: eig vs alpha} in the main text.


\section{\hl{Obtaining Fields of Line current}}
\label{app:Line current fields}
\begin{figure}
    \centering
    \includegraphics[width=0.8\linewidth]{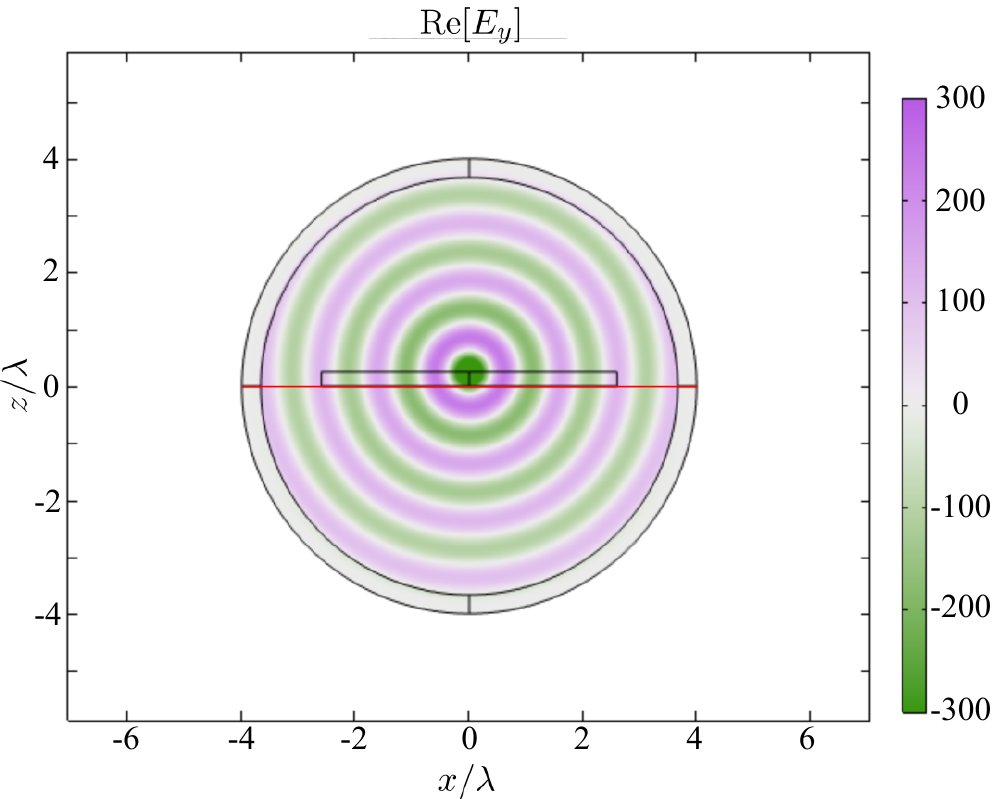}
    \caption{\hl{\textbf{Determining the incident field:}  Electric field due to a line current in free space, the central circular region containaing a line current source shifted from the origin in the $z$ direction by $\lambda/4$. The red line indicates where the fields are extracted for use as incident fields in the boundary condition given in Eq. }\ref{eq: Impedance BC}\hl{ to calculate the distribution shown in Fig.} \ref{fig:lcimp}}
    \label{fig:enter-label}
\end{figure}

\hl{In section }\ref{Sec:Lc_to_pw}\hl{ we calculated the impedance tensor necessary to reflect a TE polarized cylindrical wave into a TM polarized plane wave.  Even though an analytic solution may be available, in general we can find both incident and reflected fields numerically. In our example we modelled the line current in free space using COMSOL multiphysics}~\cite{COMSOL}\hl{, as shown in Fig. }\ref{fig:enter-label}. 

\begin{figure}
    \centering
    \includegraphics[width=1.0\linewidth]{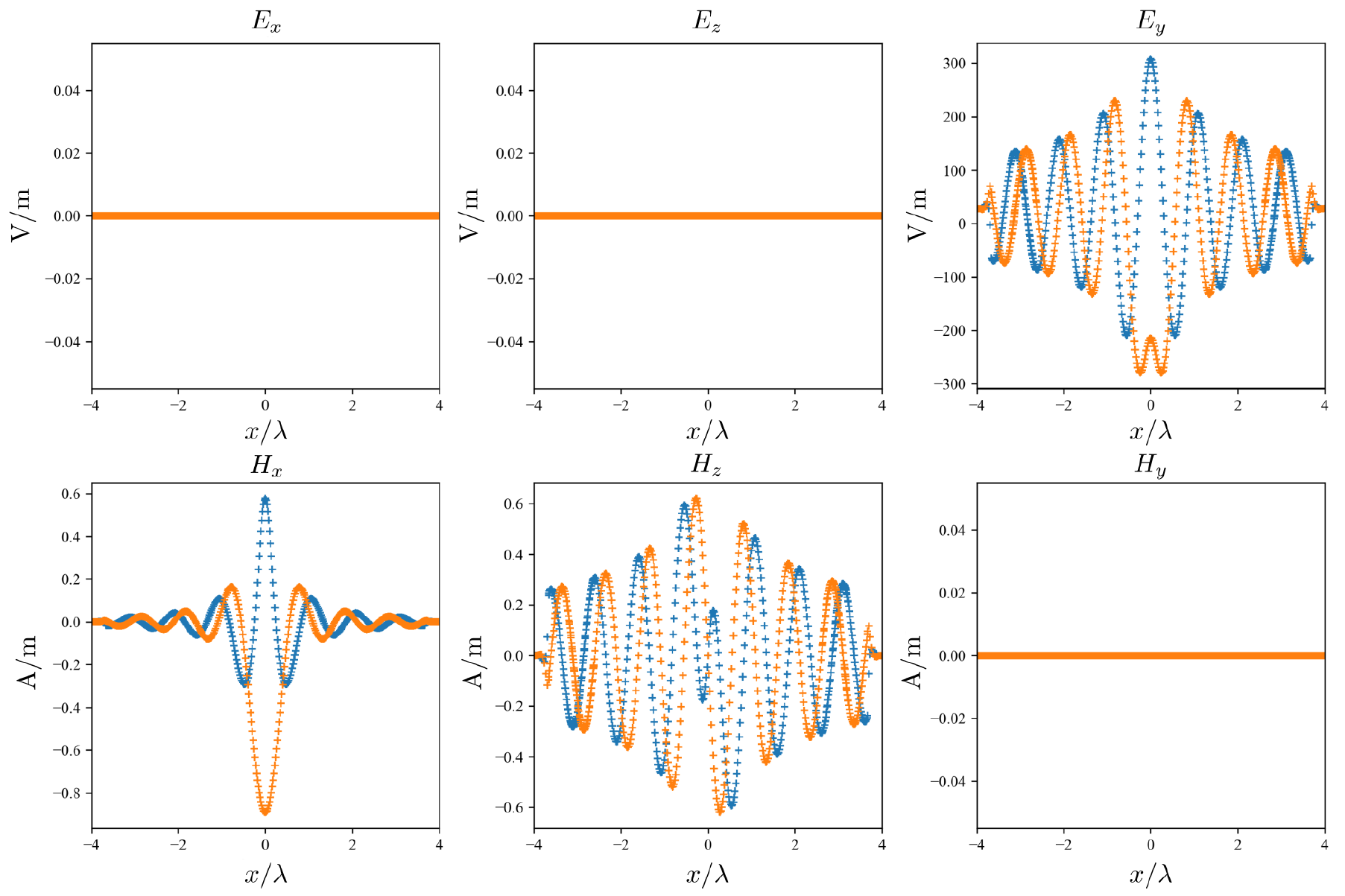}
    \caption{\hl{\textbf{Numerically calculated fields at surface:} Electric and magnetic fields evaluated at a distance $\lambda/4$ below a $\boldsymbol{\hat{y}}$ polarised line current, indicated as a red line in Fig. }\ref{fig:enter-label}}
    \label{fig:lc_fields}
\end{figure}

Tabulating the fields along the line $z=\lambda/4$ below the line current source gives us the incident electric field and magnetic fields in Eq.\ref{eq: Impedance BC}. These are shown in Fig. \ref{fig:lc_fields}.

\section{Obtaining Alpha for spatially varying fields}
\label{app:Passivealpha}
To calculate the values of the free parameter $\alpha$ appearing in Eq. \ref{eq:passive-condition-2} we write it in the form, $\alpha=C+Re^{i\phi}$ with the radius $R$,
\begin{equation*}
    \label{passR}
    R=\frac{\sqrt{|a|^2-4\Re(adc^*)+4|c|^2\Re(b)}}{2|c|}
\end{equation*}
and the centre given by,
\begin{equation*}
    \label{passC}
    C=-\frac{a^*-2dc^*}{2|c|}.
\end{equation*}
where the variables $a$, $b$, $c$, and $d$ given respectively by;
\begin{align}
    a&=\boldsymbol{\hat{h}}.\boldsymbol{\hat{h}}\nonumber\\
    b&=e_1(\boldsymbol{\hat{n}}\times\boldsymbol{\hat{h}}^{\star})\cdot\boldsymbol{\hat{h}}\nonumber\\
    c&=\sqrt{\frac{1}{4{\rm Re}(\bar{e_2})}}(\boldsymbol{\hat{h}}\cdot(\boldsymbol{\hat{n}}\times\boldsymbol{\hat{h}}^*)\nonumber\\
    d&=\sqrt{\frac{1}{4{\rm Re}(\bar{e_2})}}(\bar{e_1^*}+e_1(\boldsymbol{\hat{n}}\times\boldsymbol{\hat{h}}^*)\cdot(\boldsymbol{\hat{n}}\times\boldsymbol{\hat{h}}^*))
\end{align}
Through the parameterization in terms of an angle $\phi$ (not related to any physical angle!), a range of impedance distributions can be found, all of which are passive. For the impedance solution given in Fig. \ref{fig:lcimp} a constant value of $\phi=\pi$ is chosen.

\section{\hl{Example realisation}\label{ap:realization}}

\begin{figure}
    \centering
    \includegraphics[width=1.0\linewidth]{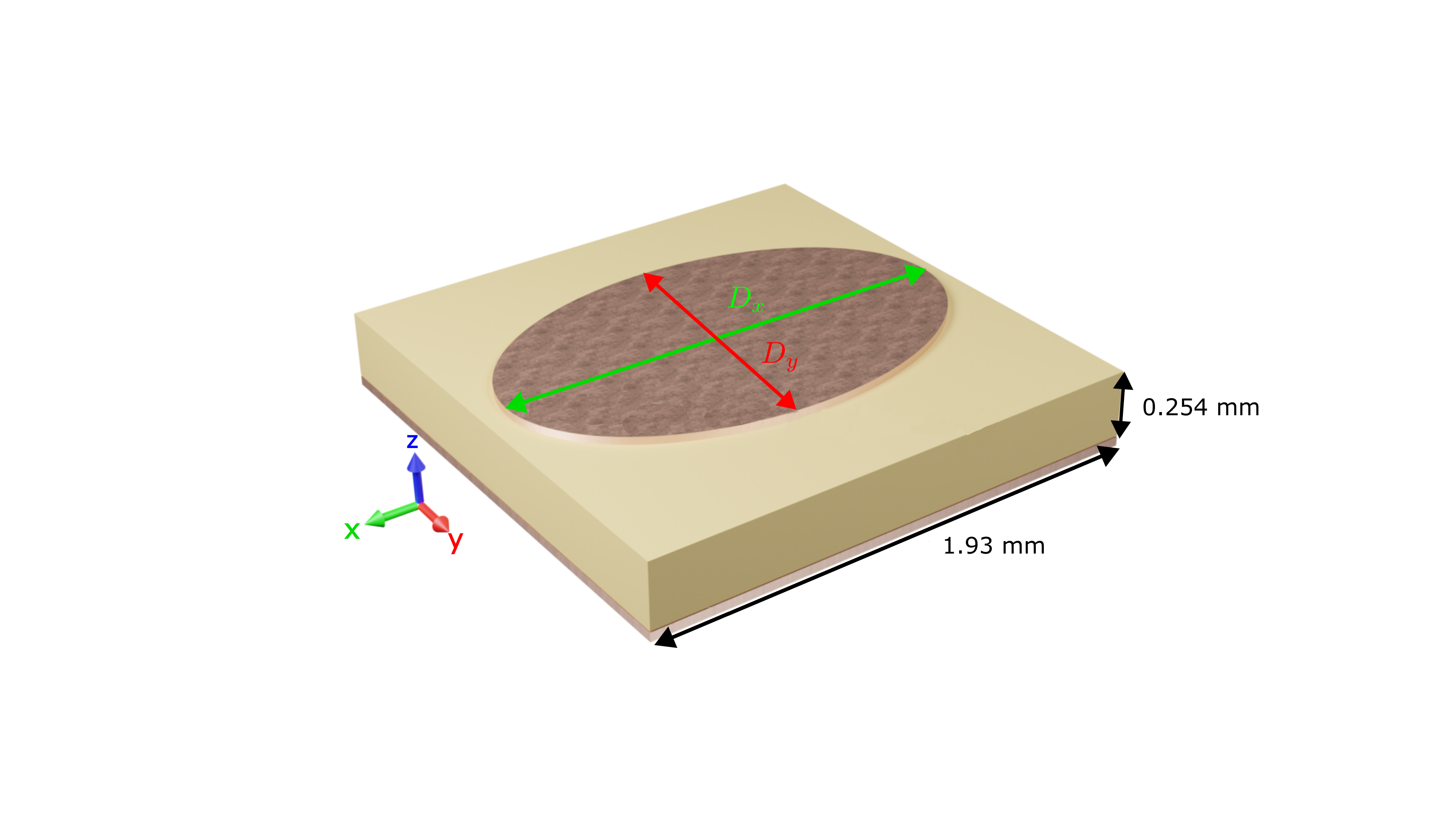}
    \caption{\hl{\textbf{Unit cell design:} Unit cell used to approximate the impedance distribution shown in Fig.} \ref{fig:diagfit}.  \hl{The top oval patch layer and bottom ground plane layer are copper. The middle dielectric layer is Ventec-VT 6710 ($\epsilon_r=10.2+0.024i$)}}
    \label{fig:unitcell}
\end{figure}

\begin{figure}
    \centering
    \includegraphics[width=0.9\linewidth]{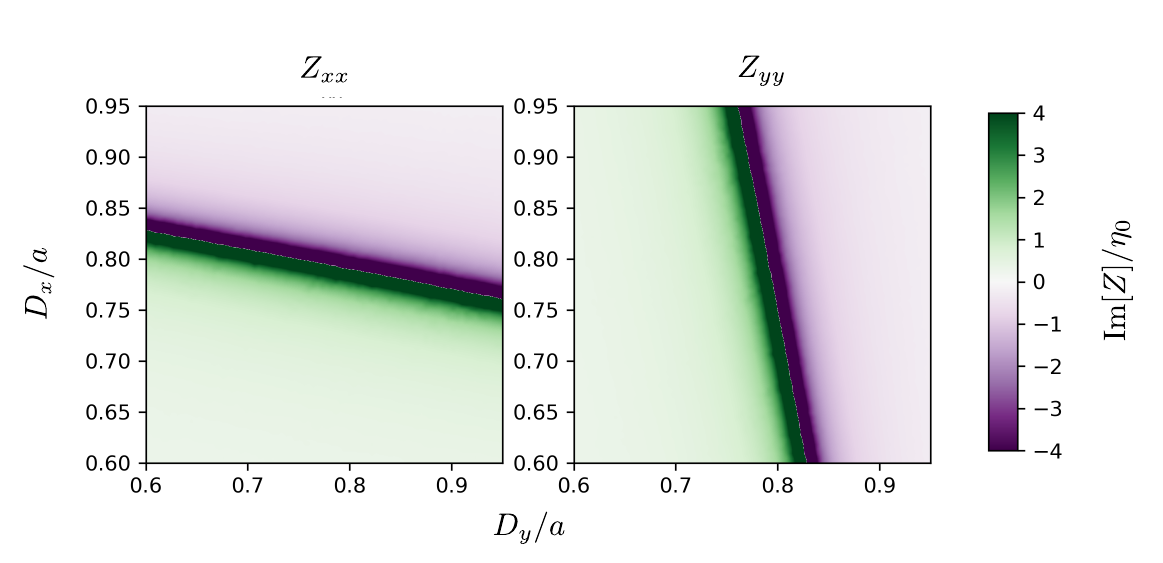}
    \caption{\hl{\textbf{Simulated impedance values}:  Impedance eigenvalues obtained from finite element modelling of the periodic array of unit-cells shown in Fig. }\ref{fig:unitcell} \hl{of appendix}~\ref{ap:realization}, \hl{for varying $x$ and $y$ diameter of the ellipse (periodicity of square lattice, $a$).  These values are derived for a plane wave at normal incident.}}\label{fig:ovalimp}
\end{figure}

\hl{To approximate the impedance distribution shown in Fig.} \ref{fig:diagfit} \hl{we used an elliptical patch on a dielectric layer on a grounded copper sheet. The parameters of the unit cell are shown in Fig.} \ref{fig:unitcell}.
\hl{By changing the major/minor axes of the elliptical patch in $x$ and $y$ (the lattice vectors of the metasurface), we can extract the impedance from either the surface fields}\cite{PatelModAn} \hl{or numerically computed reflection coefficients} \cite{smith2024method}. \hl{The surface impedance calculated from the reflection coefficients is shown in Fig.} \ref{fig:ovalimp}.  \hl{The impedance distribution over the full range of patch variations was obtained using FEM} \cite{COMSOL} \hl{using 80 steps in both the $0.6\leq D_x\leq0.9$ and $0.6\leq D_y \leq 0.9$ ($D_x$ and $D_y$ being the lengths of the elliptical patch in the $\hat{x}$ and $\hat{y}$ directions respectively). The range was then interpolated to obtain detailed functions of the surface impedance over the range to allow for fitting to the requried surface impedance distribution.}

%
%
\bibliography{main.bib}
\end{document}